\documentclass[pra,floatfix,twocolumn,showpacs,superscriptaddress]{revtex4}

\usepackage{amsmath,amssymb,amsfonts}
\usepackage{graphicx} 
\usepackage{dcolumn} 
\usepackage{bm} 

\newcommand{\ket}[1]{\left| #1 \right\rangle}
\newcommand{\bra}[1]{\left\langle #1 \right|}
\newcommand{\inner}[2]{\left\langle #1\right|\!\left. #2 \right\rangle}

\begin{document}

\title{Entangled-State Cycles of Atomic Collective-Spin States}

\author{A. Chia}
\affiliation{Centre for Quantum Dynamics, School of Biomolecular and Physical
Sciences, Griffith University, Brisbane, Queensland 4111, Australia}
\affiliation{Department of Physics, University of Auckland, Private Bag 92019, Auckland, New Zealand}

\author{A. S. Parkins}
\affiliation{Department of Physics, University of Auckland, Private Bag 92019, Auckland, New Zealand}

\date{\today}

\begin{abstract}

We study quantum trajectories of collective atomic spin states of $N$ effective two-level atoms driven with laser and cavity fields. We show that interesting ``entangled-state cycles'' arise probabilistically when the (Raman) transition rates between the two atomic levels are set equal. For odd (even) $N$, there are $(N+1)/2$ ($N/2$) possible cycles. During each cycle the $N$-qubit state switches, with each cavity photon emission, between the states $(|N/2,m\rangle\pm |N/2,-m\rangle)/\sqrt{2}$, where $|N/2,m\rangle$ is a Dicke state in a rotated collective basis. The quantum number $m$ ($>0$), which distinguishes the particular cycle, is determined by the photon counting record and varies randomly from one trajectory to the next. For even $N$ it is also possible, under the same conditions, to prepare probabilistically (but in steady state) the Dicke state $|N/2,0\rangle$, i.e., an $N$-qubit state with $N/2$ excitations, which is of particular interest in the context of multipartite entanglement. 

\end{abstract}

\maketitle
 
\section{Introduction}
\label{Intro}

The understanding that entanglement serves as a physical resource for various quantum communication or information processing protocols has motivated both experimental and theoretical studies in the preparation and quantification of entangled states \cite{Nielsen,Bouwmeester00}. 
Entanglement between two subsystems of a quantum system, i.e., bipartite entanglement, is well understood and can be directly quantified by a number of readily computable measures. Multipartite quantum systems, however, offer a richer variety of entangled states, the characterization of which is, naturally, more complicated and indeed still under active investigation (see, e.g., \cite{Dur00,Wong01,Collins02,Verstraete02,Stockton03,Toth05}). 

Of particular interest in this context are the $N$-qubit symmetric Dicke states \cite{Dicke54}; in particular, the states $|j,m\rangle$ (with $j=N/2$) satisfying
\begin{equation}
J_z|j,m\rangle = m|j,m\rangle , ~~~ {\bf J}^2|j,m\rangle = j(j+1)|j,m\rangle ,
\end{equation}
with $m\in \{ -j,-j+1,\ldots , j-1,j\}$, where $\{ J_z,{\bf J}^2\}$ are collective angular momentum operators and $(m+N/2)$ gives the total number of excitations in the system. These states are, with regards to entanglement, particularly robust against particle loss \cite{Stockton03}, and they are significant for the study and application of genuine multipartite entanglement \cite{Toth07,multipartiteE}.

Photonic systems have proven fruitful for the experimental preparation and detection of 
three-photon polarization-entangled $W$ states (Dicke states with one ``excitation'')
\cite{Kiesel03,Bourennane04,Eibl04,Mikami05,Bourennane06} and, very recently, four-photon 
polarization-entangled Dicke states with two excitations \cite{Kiesel07}.
Meanwhile, single collective atomic excitations -- atomic $W$ states -- have been produced in many-atom ensembles \cite{Kuzmich03,Eisaman04,Chaneliere05,Eisaman05,Felinto05} and with small collections of trapped ions \cite{Haeffner05}. The potentially long coherence times of ground electronic states of atoms (two of which constitute the qubit) make atomic systems particularly attractive for the storage and manipulation of entangled states, and a variety of schemes have been proposed for the preparation of more general Dicke states of trapped atoms or ions \cite{Lukin01,Unanyan03,Duan03,Zou03,Stockton04,Xiao06,Retzker07}.

Here we describe a system that adds to these proposed schemes, in that it may produce collective atomic states with multiple excitations, but also yields very interesting ``entangled-state dynamics'' and conditional quantum evolution, based upon photon detection in the output field of an optical cavity containing the atoms.
The cavity QED setup and excitation scheme that we employ have been considered previously in the context of generating, in steady state, two-qubit (i.e., two-atom) entangled mixed states of any allowed combination of purity and entanglement \cite{Clark03a}, although here we consider a specific limiting case of the operating conditions (not considered in \cite{Clark03a}) that does not in fact admit a unique steady state. 

In examining this specific case, we gain inspiration from a recent investigation of conditional quantum dynamics in a related cavity QED system involving two atoms in separate, cascaded cavities \cite{Gu06}, which also focussed on a specific limiting case of the operating conditions for an atom-entanglement scheme proposed earlier \cite{Clark03b}. In particular, we use the method of quantum trajectories based upon continuous monitoring of the cavity output field by photodetection to study the conditional quantum evolution of the collective state of $N$ atoms inside the (single) cavity. As in \cite{Gu06}, we observe manifestly distinct behaviours from one trajectory to the next; the system either evolves to a (steady) Dicke state with $N/2$ excitations without emitting any photons (for even $N$), or it executes a sustained cycle of ``switches'' between particular, well-defined superpositions of Dicke states, with a concomitant continuous output stream of photons. Furthermore, for $N>2$ more than one distinct switching cycle is possible, with the different cycles distinguished by the rate of photon emissions.
      
We start in Section \ref{QuantTrajs} with a brief review of the cavity QED configuration proposed in \cite{Clark03a}, after which we formulate our quantum trajectory model, making use of a convenient change of basis, which is motivated by the particular form of the effective Hamiltonian and jump operators. 
In Section \ref{steadystate} we focus briefly on the particular case in which the system evolves into a stable Dicke state with $N/2$ excitations.
Individual trajectory results for $N=2,3,4$ atoms, highlighting cyclic behavior of the state, are presented and discussed in Section \ref{ESC}. We also make some more general observations concerning cycles for larger $N$, before concluding in Section \ref{conclusion}.

\section{Physical system and theoretical model}
\label{QuantTrajs}

\subsection{Multiatom cavity QED system}
\label{PS&AE} 

\begin{figure}
\includegraphics[width=8.5cm]{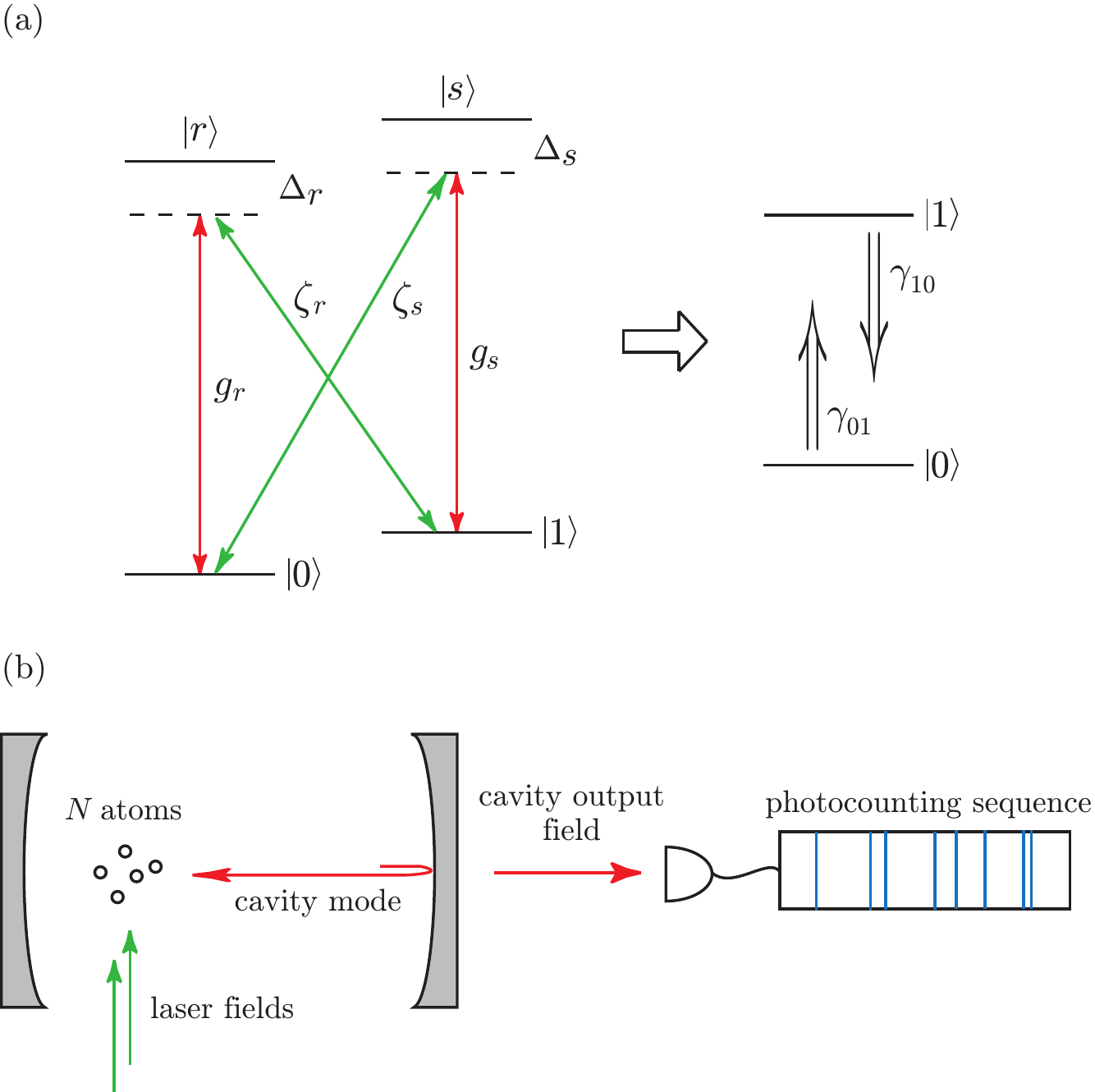} 
\caption{(a) Atomic excitation scheme and (b) physical implementation of the system studied. A single cavity field mode (coupling strength $g_{r,s}$) and two laser fields (Rabi frequencies $\zeta_{r,s}$) drive resonant Raman transitions between the atomic ground states $\ket{0}$ and $\ket{1}$. Note that an additional laser or magnetic field may be required in the excitation scheme to provide an independent relative energy shift between the ground states $\ket{0}$ and $\ket{1}$ (see \cite{Clark03a}). Note also that $\ket{r}$ and $\ket{s}$ may be a single level, provided the two Raman channels remain distinct from each other (which would require that the two ground states are nondegenerate in energy). For large detunings $\Delta_r$ and $\Delta_s$, and for $\kappa$, the cavity field decay rate, much larger than the Raman transition rates, the model may be reduced to that of an effective two-level atom with characteristic transition rates $\gamma_{01}$ and $\gamma_{10}$. }
\label{PhysSys}
\end{figure}

We consider a system of $N$ four-level atoms trapped inside a high finesse optical cavity. The atomic excitation scheme and physical implementation are depicted in Fig.~\ref{PhysSys}. A single quantized cavity mode couples to the $\ket{0} \leftrightarrow \ket{r}$ and $\ket{1} \leftrightarrow \ket{s}$ transitions in each atom with coupling strengths $g_r$ and $g_s$, respectively. Auxiliary laser fields couple to the $\ket{0} \leftrightarrow \ket{s}$ and $\ket{1} \leftrightarrow \ket{r}$ transitions with Rabi frequencies $\zeta_r$ and $\zeta_s$, respectively. Together, the cavity and laser fields drive (resonant) Raman transitions between the stable atomic ground states $\ket{0}$ and $\ket{1}$. 
Such an excitation scheme could be achieved, for example, by using orthogonal, linearly-polarized laser and cavity fields on an $F=1/2\leftrightarrow F'=1/2$ atomic transition (as in ${}^6\textrm{Li}$), or on an $F=1\leftrightarrow F'=0$ transition (as in ${}^{87}\textrm{Rb}$) with an applied magnetic field, as described in \cite{Dimer07}.

Finally, we note that the spacings between the atoms are assumed to be sufficiently large that direct dipole-dipole interactions can be neglected. We also assume that each atom is held tightly in position by a confining trap, so that the atomic motion can be omitted from the analysis.

%


\subsection{Reduced $N$-atom master equation}
\label{JxDescription}

With the assumption of large detunings $\Delta_r$ and $\Delta_s$, the atomic excited states can be adiabatically eliminated and atomic spontaneous emission neglected. The coherent dynamics is then characterized by Raman transition rates and energy level shifts, which we subsequently assume to be much smaller than the cavity field decay rate, $\kappa$. This enables us to adiabatically eliminate the cavity field to derive a master equation that describes the atomic dynamics alone. Under ``resonant'' conditions (i.e., light shifts of the atomic ground states are balanced, such that Raman resonance is maintained; see \cite{Clark03a} for details), this master equation can be written in the form
\begin{eqnarray}  
\label{MasterEq1} 
   \frac{d\rho}{dt} = 2X\rho X^\dag -X^\dag X\rho - \rho X^\dag X ,
\end{eqnarray}     
where $\rho$ is the atomic density matrix, 
\begin{eqnarray}  
X = \sqrt{\gamma_{01}}\, J_+ + \sqrt{\gamma_{10}}\, J_-
\end{eqnarray}     
with $\gamma_{01}=\left|\zeta_sg_s/2\Delta_s\right|^2/\kappa$,
$\gamma_{10}=\left|\zeta_rg_r/2\Delta_r\right|^2/\kappa$, and
\begin{eqnarray}  
J_+=\sum_{i=1}^N \ket{1}\bra{0}_i \, , ~~~ J_- = (J_+)^\dag .
\end{eqnarray}     
For $\gamma_{01}\neq\gamma_{10}$, the dynamics of (\ref{MasterEq1}) produces unique steady states that are non-maximally-entangled (pure or mixed) collective spin states \cite{Palma89,Agarwal89,Clark03a}. However, for the remainder of this paper we consider the case 
\begin{eqnarray}
   \gamma_{01} = \gamma_{10}  \,,
\label{resonance}   
\end{eqnarray}
for which (\ref{MasterEq1}) does not admit a unique steady state (mathematically, the system of linear equations for the density matrix elements possesses a second zero eigenvalue when $\gamma_{01} = \gamma_{10}$). Scaling time by $(4\gamma_{01})^{-1}$, the master equation we thus focus on takes the form
\begin{eqnarray}  
\label{MasterEq2a} 
   \frac{d\rho}{dt} = 2J_{x}\rho J_{x} - J^{2}_{x}\,\rho - \rho J^{2}_{x} ,
\end{eqnarray}     
where $J_{x}=(J_++J_-)/2$ is the $x$-component of the total pseudo-spin $\mathbf{J}$ of all $N$ qubits.
Note that since $\mathbf{J}^2$ commutes with $J_x$, the dynamics described by (\ref{MasterEq2a})
conserves the magnitude of the total spin angular momentum (which we will take to be the maximal value in this work) \cite{SpE}.


\subsection{Quantum trajectories in the collective-spin representation}

\subsubsection{Change of basis}

We are interested in the time-dependent behavior of the atomic state subject to dynamics as described by (\ref{MasterEq2a}) and, as depicted in Fig.~\ref{PhysSys}(b), in photon counting measurements on the cavity output field. The method of quantum trajectories (see, e.g., \cite{Carmichael89,Carmichael93,Dalibard92,Dum92}), summarized briefly in the Appendix, lends itself nicely to this task.

In this approach, and with photon counting measurements in mind, we rewrite (\ref{MasterEq2a}) in the form
\begin{eqnarray}  
\label{MasterEq2b} 
   \frac{d\rho}{dt} = -i\mathcal{H}_{\rm eff}\rho + i\rho \mathcal{H}_{\rm eff}^\dag + 2C\rho C^\dag ,
\end{eqnarray}     
and identify a non-Hermitian effective Hamiltonian, $\mathcal{H}_{\rm eff}$, and collapse operator, $C$, as
\begin{eqnarray}
\label{MasterEq2c}
   \mathcal{H}_{\rm eff}=-iJ^{2}_{x}\,, \quad C=J_{x} \,.
\end{eqnarray}
In the trajectory model, free evolution of the atomic wave function, subject to $\mathcal{H}_{\rm eff}$, is interspersed, at random times, with collapses (or quantum jumps) produced by the action of $C$ on the wave function. These collapses can be identified with photon detections in the cavity output field and the resulting wave function trajectory describes the state of the atomic spin system conditioned upon the sequence of photon detections.

Given the form of both $\mathcal{H}_{\rm eff}$ and $C$, a particularly convenient basis for our study will be the basis of eigenstates of the $x$-component of the total spin $\mathbf{J}$; more precisely, simultaneous eigenstates of $\mathbf{J}^{2}$ and $J_x$. In the more familiar representation of simultaneous eigenstates of $\mathbf{J}^{2}$ and $J_{z}$, which we denote by $|j,m\rangle_z$, the allowed values of the $J_z$-component are $m=-j,-j+1,\ldots,j$, where, for a system of $N$ effective spin-1/2 particles, as we consider here, $j$ may take one of the values $\{0, 1/2, 1,\ldots, N/2\}$. 
So, for example, for $N=2$ atoms and $m=0$,
\begin{subequations}
\begin{eqnarray}
\label{z_basis=bare_states_a}
   \ket{0,0}_z &=& \frac{1}{\sqrt{2}}\left( \ket{10}_z - \ket{01}_z \right) \,, \\
   \ket{1,0}_z &=& \frac{1}{\sqrt{2}}\left( \ket{10}_z + \ket{01}_z \right) \,,
\end{eqnarray}
while for $m=\pm1$ (both spins ``up'' or ``down'')  
\begin{eqnarray}
\label{2Q_z_basis=bare_states_b}
   \ket{1,-1}_z = |00\rangle_z \,, \quad \ket{1,1}_z = \ket{11}_z \, ,
\end{eqnarray}
\end{subequations}
where $\ket{ij}_z$ $(i,j=0,1)$ denotes the state with atom 1 in state $\ket{i}_z$ and atom 2 in state $\ket{j}_z$.
   
To obtain the eigenstates of $J_{x}$, we perform the appropriate set of Euler rotations on the coordinate system. In particular, the $J_x$-eigenstates are related to the $J_z$-eigenstates by a unitary rotation operator with the property $D^{-1}(\beta)=D(-\beta)$:
\begin{eqnarray}
\label{z_basis=x_basis}
   \ket{j,n}_z = \sum^{j}_{m=-j} D^{(j)}_{m,n}\! \left(-\frac{\pi}{2}\right) \ket{j,m}_x \,.
\end{eqnarray}
Here, $D^{(j)}_{m,n}(\beta)$ are the matrix elements of the rotation operator given by (sometimes referred to as Wigner's formula) \cite{Sakurai}
\begin{widetext}
\begin{eqnarray}
\label{WignersFormula}
   D^{(j)}_{m,n}\left(\beta\right) = \sum^{j+m}_{k=0} (-1)^{k-n+m}\frac{\sqrt{(j+n)!\,(j-n)!\,(j+m)!\,(j-m)!}}{k!\,(j+n-k)!\,(j-m-k)!\,(k+m-n)!}                                            \left(\cos\frac{\beta}{2}\right)^{2j-2k+n-m} \left(\sin\frac{\beta}{2}\right)^{2k-n+m} \,. 
\end{eqnarray}
\end{widetext}
Note that the sum in (\ref{WignersFormula}) is only over terms with non-negative arguments of the factorials. 

Finally, we will confine our study to evolution from the initial state
\begin{eqnarray}
   \ket{\psi_N(0)} = \ket{N/2,-N/2}_z = |0^{\otimes N}\rangle_{z} \,, 
\end{eqnarray}
i.e., $j=N/2$ (the maximal value) and all spins ``down''. Since $j$ is conserved by the evolution, only $N+1$ linearly independent basis states are required to describe the state of the $N$-atom system.

\subsubsection{Jumps versus free evolution: general behavior}

In the basis of $J_x$-eigenstates it is very straightforward and instructive to calculate the general behavior of the spin wave function either undergoing free evolution subject to $\mathcal{H}_{\rm eff}$, or being ``collapsed'' under the action of the jump operator $C=J_x$. In particular, consider first free evolution from the initial state $\ket{\psi_N(0)}=\ket{N/2,-N/2}_{z}$ up to some time $t$, without any collapses having occurred. For even $N$, the (unnormalized) state $|\bar{\psi}_N(t)\rangle$ can be written as
\begin{eqnarray}
\label{TimeEvo}
   |\bar{\psi}_N(t)\rangle &=& e^{-i\mathcal{H}_{\rm eff}t} \ket{\psi_N(0)} \nonumber \\[0.2cm]
                   &=& D^{(N/2)}_{0,-N/2}\left(-\frac{\pi}{2}\right) \ket{N/2,0}_{x} \nonumber \\
                    && + \sum^{N/2}_{\substack{m=-N/2 \\ m\ne0}} D^{(N/2)}_{m,-N/2}\left(-\frac{\pi}{2}\right) \:e^{\!-m^{2}t} \ket{N/2,m}_x . \nonumber\\
\end{eqnarray}
Noting that $D^{(N/2)}_{m,-N/2}\left(-\frac{\pi}{2}\right)=D^{(N/2)}_{-m,-N/2}\left(-\frac{\pi}{2}\right)$, it is useful to define
\begin{eqnarray}
\label{esc}
  |\chi^{\pm}_{N}(m)\rangle = \frac{1}{\sqrt{2}}\:(\,\ket{N/2,m}_x \pm \ket{N/2,-m}_x \,) \,,
\end{eqnarray} 
and rewrite (\ref{TimeEvo}) as
\begin{eqnarray}
\label{evo}
   |\bar{\psi}_{N}(t)\rangle &=& D^{(N/2)}_{0,-N/2}\left(-\frac{\pi}{2}\right) \ket{N/2,0}_{x} \nonumber
   \\
   && + \sum_{m=1}^{N/2} c_m e^{-m^{2}t} \;|\chi^{+}_{N}(m)\rangle ,
\end{eqnarray} 
where $c_m=\sqrt{2}D^{(N/2)}_{m,-N/2}\left(-\frac{\pi}{2}\right)$. 
This result shows explicitly that the $m=0$ component of the spin wave function in the $J_x$-eigenstate basis is unaffected by the free evolution, while the amplitudes of the $m\neq0$ components decay exponentially with time. 

Conversely, the $m=0$ component is completely removed by the action of the jump operator, i.e., $J_x\ket{N/2,0}_{x}=0$, while
\begin{eqnarray}
\label{jumps}
  J_{x} \ket{\chi^{\pm}_{N}(m)} = m \ket{\chi^{\mp}_{N}(m)} .
\end{eqnarray}   

For odd $N$, there is no $m=0$ component and the (unnormalized) wave function takes the form
\begin{eqnarray}
\label{evo_odd}
   |\bar{\psi}_{N}(t)\rangle = \sum_{m=1/2}^{N/2} c_m e^{-m^{2}t} \;|\chi^{+}_{N}(m)\rangle ,
\end{eqnarray} 
after a time $t$ of free evolution (from the initial state $\ket{\psi_N(0)}=\ket{N/2,-N/2}_{z}$). Hence, all components of the wave function have amplitudes that decay exponentially.

Now, in the trajectory formalism, the square of the norm of the wave function, $\langle \bar{\psi}_{N}(t)|\bar{\psi}_{N}(t) \rangle\leq 1$, can be identified as the probability for a collapse (i.e., a photon detection) {\it not to have happened} \cite{Meystre&Sargent}. From this identification and our observations above, it follows that the system under investigation can exhibit very distinct behaviors, as we detail in the following two sections.


\section{Steady state preparation of the Dicke state $\ket{N/2,0}_{x}$}
\label{steadystate}

From (\ref{evo}), it follows that for an even number $N$ of atoms there is a finite probability, given by
\begin{eqnarray}
P_N(m=0) = 
\left| D^{(N/2)}_{0,-N/2}\left(-\frac{\pi}{2}\right) \right|^2 = \frac{N!}{2^N\left[ (N/2)!\right]^2} \, ,
\label{P_Nm0}
\end{eqnarray}
that no collapse ever occurs and, hence, that the system evolves to a stable steady state given by $\ket{N/2,0}_{x}$. This is a Dicke state, in the basis of $J_x$-eigenstates, with $N/2$ excitations, which, as mentioned in the introduction, is of particular interest in the context of multipartite quantum entanglement. In fact, for our system a necessary and sufficient condition for genuine $N$-partite entanglement is \cite{Toth07,Korbicz05}
\begin{eqnarray}
\langle J_x^2 \rangle < \frac{N}{4} \, ,
\end{eqnarray}
which is ``optimally'' satisfied for the state $\ket{N/2,0}_{x}$, for which $\langle J_x^2 \rangle=0$.
Note that with a standard unitary rotation, implemented for example using coherent laser Raman pulses, this state can be transformed into the corresponding state $\ket{N/2,0}_{z}$ in the (conventional) $J_z$-eigenstate basis.

\begin{figure}
\includegraphics[width=8cm]{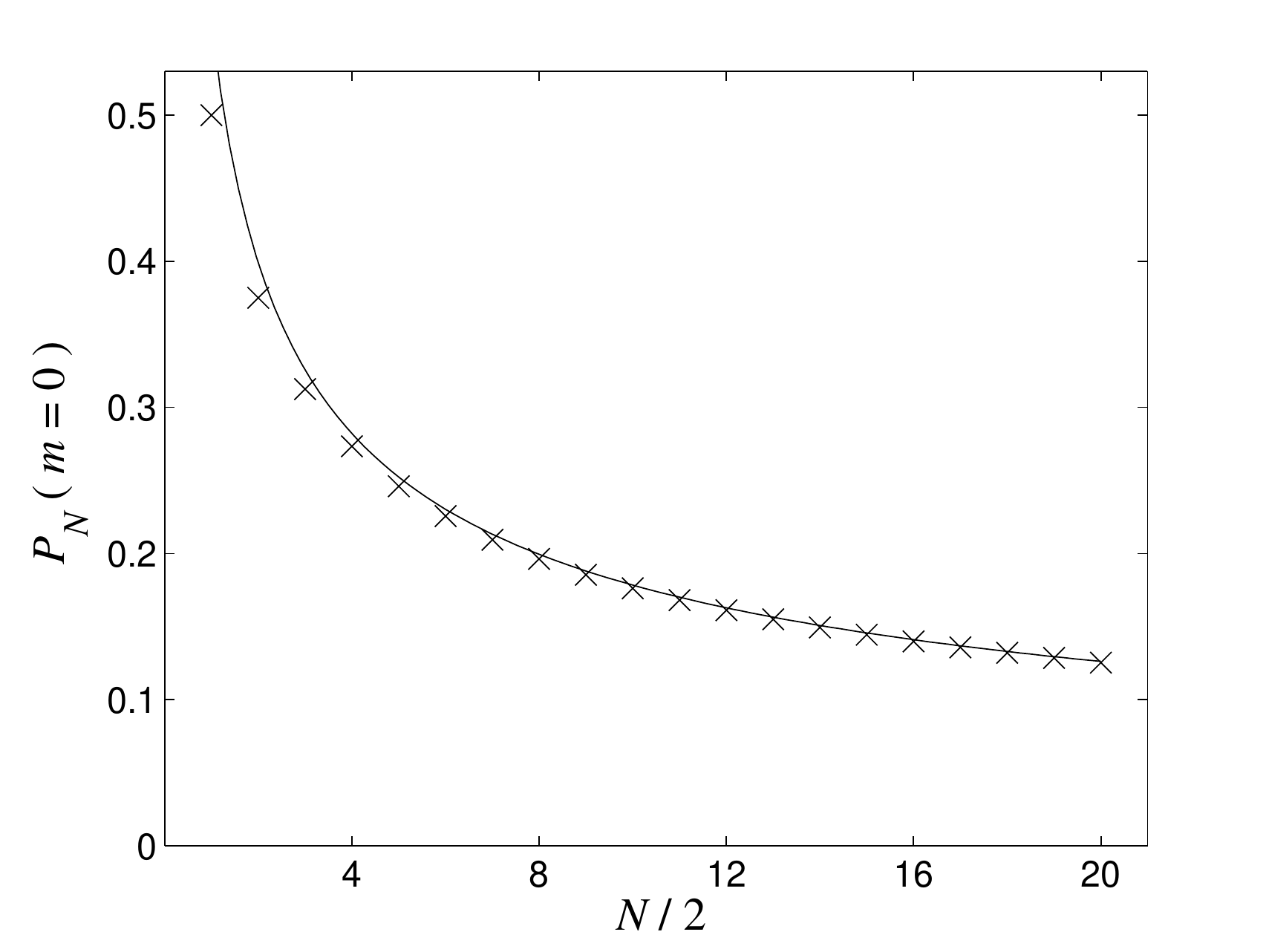} 
\caption{Probability $P_N(m=0)$ for preparation of the Dicke state $\ket{N/2,0}_{x}$ versus $N/2$ ($N$ even). The crosses are calculated from Eq.~(\ref{P_Nm0}). The solid line is the curve $[2/(\pi N)]^{1/2}$, which approximates $P_N(m=0)$ for $N\gg 1$.}
\label{Pjmx0}
\end{figure}

The probability $P_N(m=0)$ is plotted in Fig.~\ref{Pjmx0} as a function of $N$ and maintains a significant value even for quite large $N$. In fact, using Stirling's approximation ($n!\simeq n^ne^n\sqrt{2\pi n}$ for large $n$), one can show that for $N\gg1$,
\begin{eqnarray}
P_N(m=0) \simeq \sqrt{\frac{2}{\pi N}} \, ,
\end{eqnarray}
which describes the slow decrease with $N$, as illustrated in Fig.~\ref{Pjmx0}.

It must be emphasized that in the case that the state $\ket{N/2,0}_{x}$ is prepared, no photons are emitted from the cavity.
With a probability $1-P_N(m=0)$ there {\it is} a collapse, or photon detection. This in fact initiates a sustained sequence of collapses (and photon detections), since the amplitudes of all of the remaining ($m\neq0$) components of the wave function decay exponentially under free evolution, i.e., the norm of the wave function is now guaranteed to decay towards zero, meaning that a collapse is also guaranteed to occur. Hence, the system exhibits a distinctly bimodal behavior: either no photons are detected, ``signalling'' preparation of the state $\ket{N/2,0}_{x}$, or a continuous stream of photons is detected. While this stream of photons indicates failure in the preparation of a steady state, it also signals the initiation of some remarkable dynamics of the system, which we label {\it entangled state cycles} and examine in the next section.

%


\section{Entangled-State Cycles}   
\label{ESC}  

As described above, a first photon detection initiates a persistent sequence of photon detections and the relevant atomic spin states involved in the dynamics from then on are the states $|\chi^{\pm}_{N}(m)\rangle$. Note that for odd $N$, a first photon detection and subsequent persistent sequence is guaranteed to occur.

The rate of photon detections is proportional to the expectation value of the operator $C^\dag C=J_x^2$, and hence to the value $m^2$. The photon detection configuration we consider can therefore be interpreted as providing a measurement of $m^2$, so we can expect each trajectory to ``converge'' towards a particular pair of states, $|\chi^{\pm}_{N}(m)\rangle$ (or possibly to the state $\ket{N/2,0}_{x}$ for even $N$), conditioned upon the measurement record. For example, from (\ref{jumps}) we see that collapses increase the amplitude $\inner{\chi^{\pm}_{N}(m+1)}{\psi_{N}(t)}$ relative to $\inner{\chi^{\pm}_{N}(m)}{\psi_{N}(t)}$, while from (\ref{evo}), we see that free evolution decreases $\inner{\chi^{\pm}_{N}(m+1)}{\psi_{N}(t)}$ exponentially relative to $\inner{\chi^{\pm}_{N}(m)}{\psi_{N}(t)}$. 
Hence, within a given time interval, the more photon detections that occur, the more probable it is for the state to converge towards states $|\chi^{\pm}_{N}(m)\rangle$ with larger $m$ (for which, of course, a higher photon count rate is also expected). In contrast, the longer $|\psi_{N}(t)\rangle$ evolves without a collapse the more likely it is for the state to converge towards states $|\chi^{\pm}_{N}(m)\rangle$ with lower values of $m$. 

Which state or pair of states ($\ket{N/2,0}_{x}$ or $|\chi^{\pm}_{N}(m)\rangle$) is ``selected'' in a particular trajectory is random. In the case of $|\chi^{\pm}_{N}(m)\rangle$, the sequence of photon detections corresponds to a sequence of jumps, described by (\ref{jumps}), between the states $|\chi^{+}_{N}(m)\rangle$ and $|\chi^{-}_{N}(m)\rangle$. These states are also very interesting entangled states, and we label the system's behavior in this instance as an {\em entangled-state cycle}.  
We now look at specific examples of entangled state cycles for $N=2,3,4$. 
                                                  

\subsection{Entangled-state cycles for \textsl{N}$=$ 2}  
\label{N=2}
\begin{figure}
\includegraphics[width=8cm]{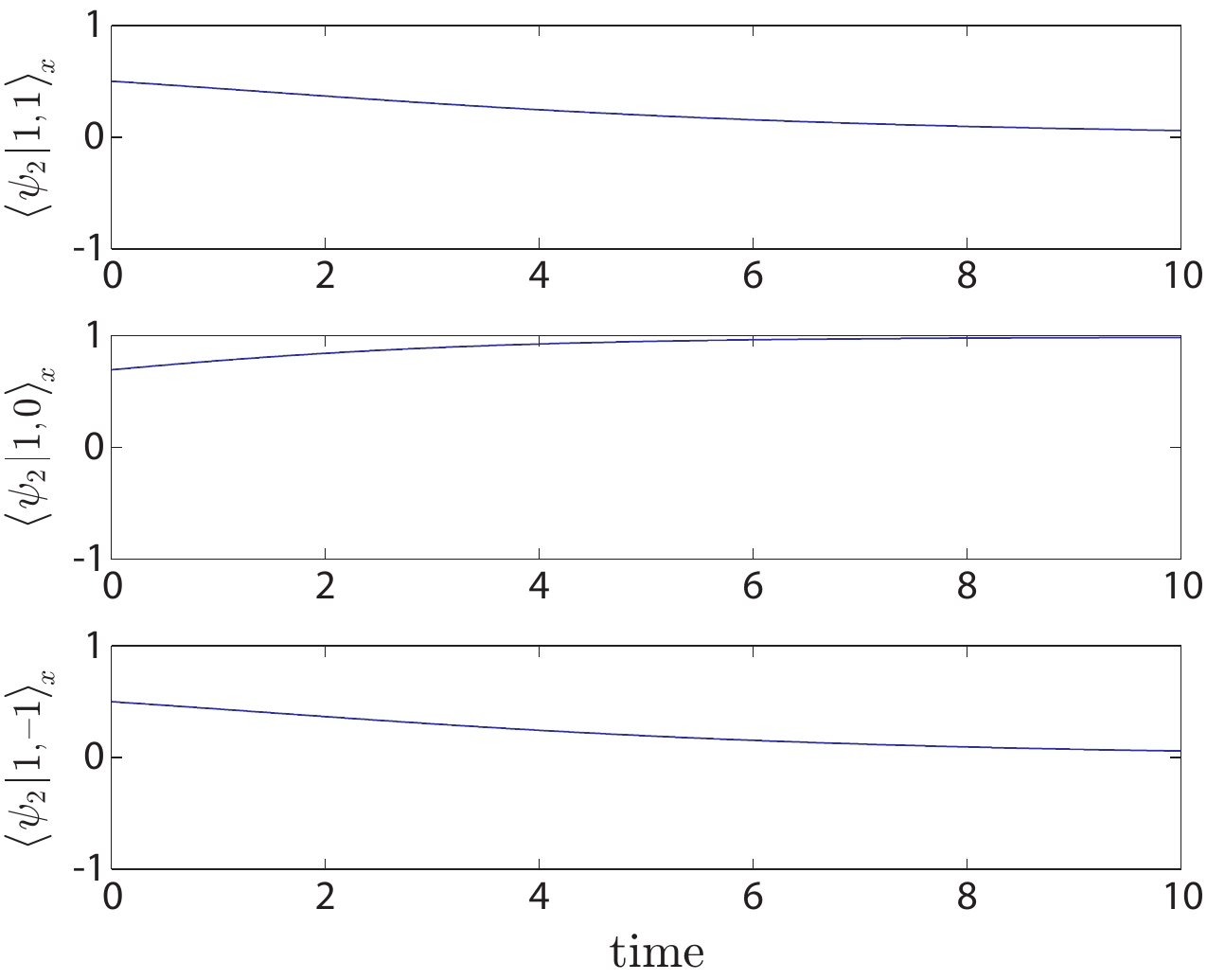} 
\caption{Example trajectory for the initial state $\ket{1,-1}_{z}=\ket{00}_z$. Time is measured in units of $(4\gamma_{10})^{-1}$. No collapses occur and the system is eventually projected into the $\ket{1,0}_x$ eigenstate.}
\label{2QPic1}
\end{figure}

Consider free evolution of the initial state 
\begin{eqnarray}
\ket{\psi_{2}(0)} = \ket{1,-1}_{z} = \frac{1}{\sqrt{2}} \left( \ket{1,0}_x + |\chi^{+}_{2}(1)\rangle \right) 
\end{eqnarray}
from $t=0$ up to time $t$, without any collapses having occurred. 
The normalized wave function is given by
\begin{eqnarray}
\label{2Q_psi(t)}
   \ket{\psi_{2}(t)} = \frac{e^{-i\mathcal{H}_{\rm eff}t}\ket{\psi_{2}(0)}}{\left\| e^{-i\mathcal{H}_{\rm eff}t}\ket{\psi_{2}(0)} \right\|} 
                     = \frac{\ket{1,0}_x + \:e^{-t}|\chi^{+}_{2}(1)\rangle}{\sqrt{(1+e^{-2t})}} \,. \nonumber\\   
\end{eqnarray} 
This illustrates very clearly that as time progresses and no collapses occur (i.e., no photons are detected), the more likely it is for the system to be in the state $\ket{1,0}_x$. Such a no-collapse trajectory (an example of which is shown in Fig.~\ref{2QPic1}) occurs with probability
\begin{eqnarray}
\label{2QbitssProb}
  P_2(m=0) =  |\langle\psi_2(0)|1,0\rangle_x|^{2} = \frac{1}{2} \,,
\end{eqnarray}
i.e., in 50\% of trials the system is prepared in the two-qubit maximally-entangled state 
\begin{eqnarray}
\label{2QbitS}
   \ket{1,0}_x = \frac{1}{\sqrt{2}}\left( \ket{1,-1}_z - \ket{1,1}_z \right) = \frac{1}{\sqrt{2}}\left( \ket{00}_z - \ket{11}_z \right) \,. \nonumber \\
\end{eqnarray}

\begin{figure}
\includegraphics[width=8cm]{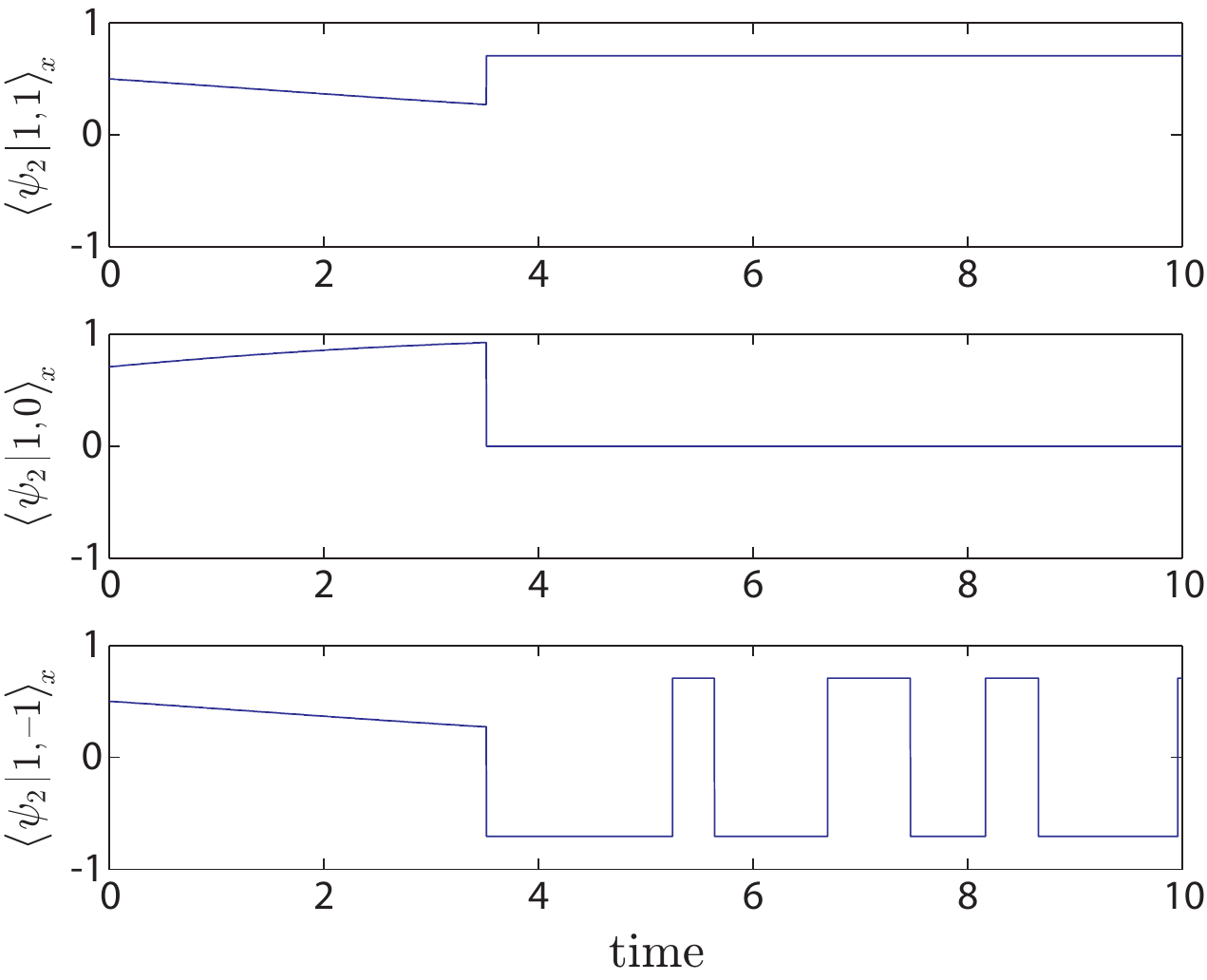} 
\caption{Example trajectory for the initial state $\ket{1,-1}_{z}=\ket{00}_z$. The system settles into an entangled-state cycle after the first collapse.}
\label{2QPic2}
\end{figure}

When a collapse does occur and a photon is detected, the system state is projected onto   
\begin{eqnarray}
\label{2QbitC}
   \ket{\chi^{-}_{2}(1)} = \frac{C\ket{\psi_2(t)}}{\left\|C\ket{\psi_2(t)}\right\|} = \frac{1}{\sqrt{2}} \left( \ket{1,1}_x - \ket{1,-1}_x \right) \, .
                             \nonumber \\                         
\end{eqnarray}
With the $m=0$ component removed, subsequent time evolution is guaranteed to lead to further collapses and a persistent cycle is established between the two states $\ket{\chi^{\pm}_{2}(1)}$. Such a cycle is illustrated in Fig.~\ref{2QPic2}, where the abrupt changes in the sign of the amplitude $\langle\psi_2|1,-1\rangle_x$ indicate jumps between the states $\ket{\chi^{\pm}_{2}(1)}$.

It is interesting to consider the compositions of the states $\ket{\chi^{\pm}_{2}(1)}$ in the original 
$J_z$-eigenstate basis; in particular,
\begin{subequations}
\begin{eqnarray}
\label{2QbitC_bare1}
   \ket{\chi^{-}_{2}(1)} &=& \ket{1,0}_z = \frac{1}{\sqrt{2}} \left( \ket{01}_z + \ket{10}_z \right) \,, 
\end{eqnarray}   
and
\begin{eqnarray}
\label{2QbitC_bare2}
   \ket{\chi^{+}_{2}(1)} &=& \frac{1}{\sqrt{2}}\left( \ket{1,1}_z + \ket{1,-1}_z \right) = \frac{1}{\sqrt{2}}\left( \ket{00}_z + \ket{11}_z \right) .                                 \nonumber \\
\end{eqnarray}     
\end{subequations}
So, when a cycle is established, the states involved are also maximally-entangled states of the two qubits.
%


\subsection{Entangled-state cycles for \textsl{N}=3}  
\label{N=3}

Considering the initial state
\begin{eqnarray}
|\psi_{3}(0)\rangle &=& \ket{3/2,-3/2}_z \nonumber
\\
&=& \sqrt{\frac{3}{4}}\ket{\chi^{+}_{3}\!\left(\tfrac{1}{2}\right)} 
                                     + \frac{1}{\sqrt{4}}\ket{\chi^{+}_{3}\!\left(\tfrac{3}{2}\right)} , 
\end{eqnarray}  
we see that there is no $m=0$ component and hence an entangled state cycle always results. In particular, a first collapse is guaranteed, as exemplified by the wave function after time $t$ of free evolution (ignoring normalization for simplicity, i.e., Eq.~(\ref{evo_odd}) for $N=3$),
\begin{eqnarray}
    |\bar{\psi}_{3}(t)\rangle = \sqrt{\frac{3}{4}}\:e^{-t/4} \ket{\chi^{+}_{3}\!\left(\tfrac{1}{2}\right)} 
                                     + \frac{1}{\sqrt{4}}\:e^{-9t/4}\ket{\chi^{+}_{3}\!\left(\tfrac{3}{2}\right)}\! .
\end{eqnarray}  

\begin{figure}
\includegraphics[width=8cm]{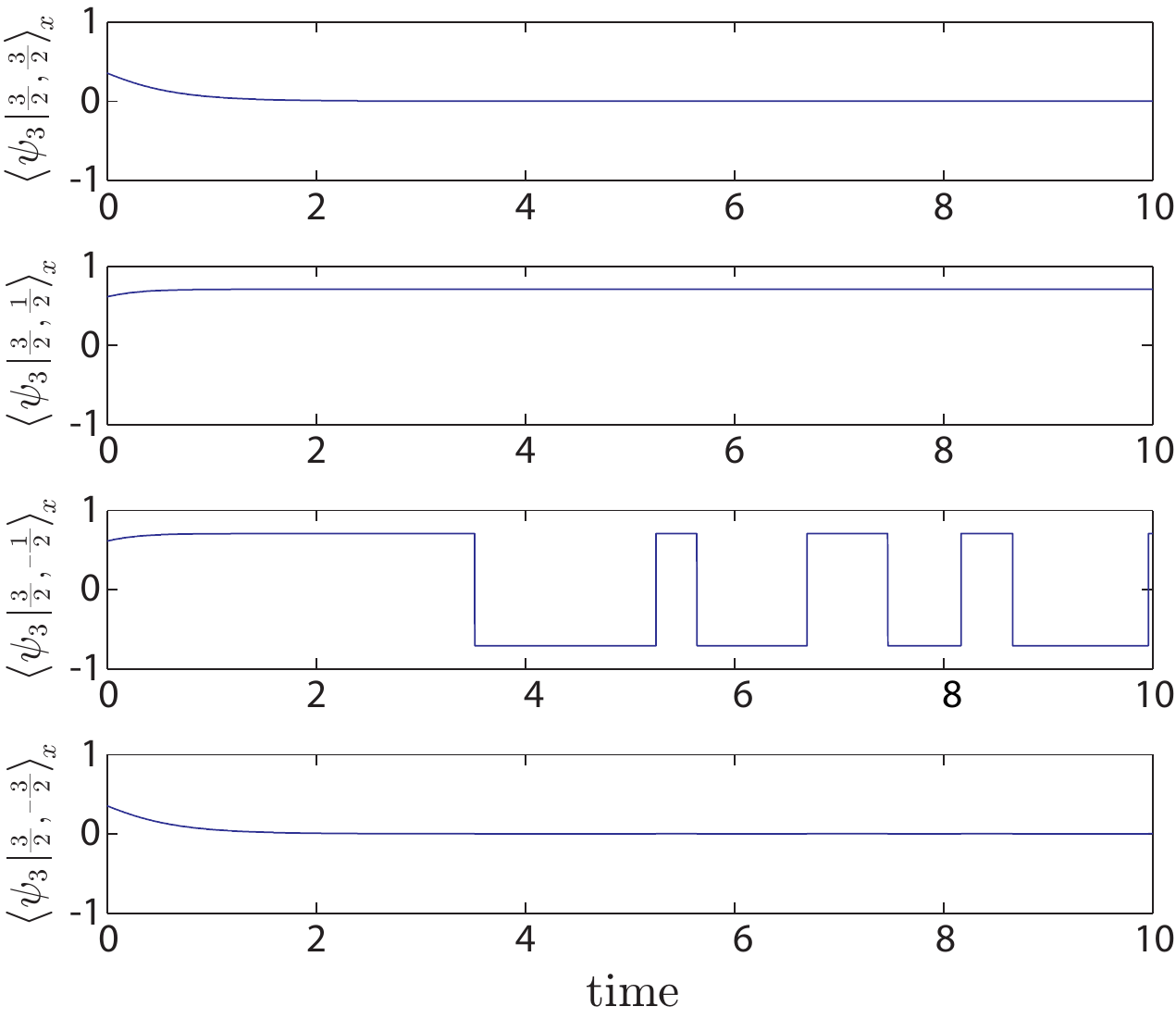} 
\caption{Example trajectory for the initial state $\ket{3/2,-3/2}_{z}=\ket{000}_z$. The $m=\pm3/2$ components decay to a negligible magnitude by the time of the first collapse and the system evolves into the cycle between the states $\ket{\chi^{\pm}_{3}(1/2)}$.}
\label{3QPic1}
\end{figure}

With probability 
\begin{eqnarray}
   |\langle\chi^{+}_{3}\!\left(\tfrac{1}{2}\right)|\psi_{3}(0)\rangle|^{2} = \frac{3}{4} \,,
\end{eqnarray} 
the system evolves to an entangled-state cycle between the states $\ket{\chi^{\pm}_{3}(1/2)}$. A sample trajectory of this cycle is shown in Fig.~\ref{3QPic1}. In this particular example, by the time of the first collapse, or photon detection, the components of $\ket{3/2,\pm3/2}_x$ have become negligibly small. 

\begin{figure}
\includegraphics[width=8cm]{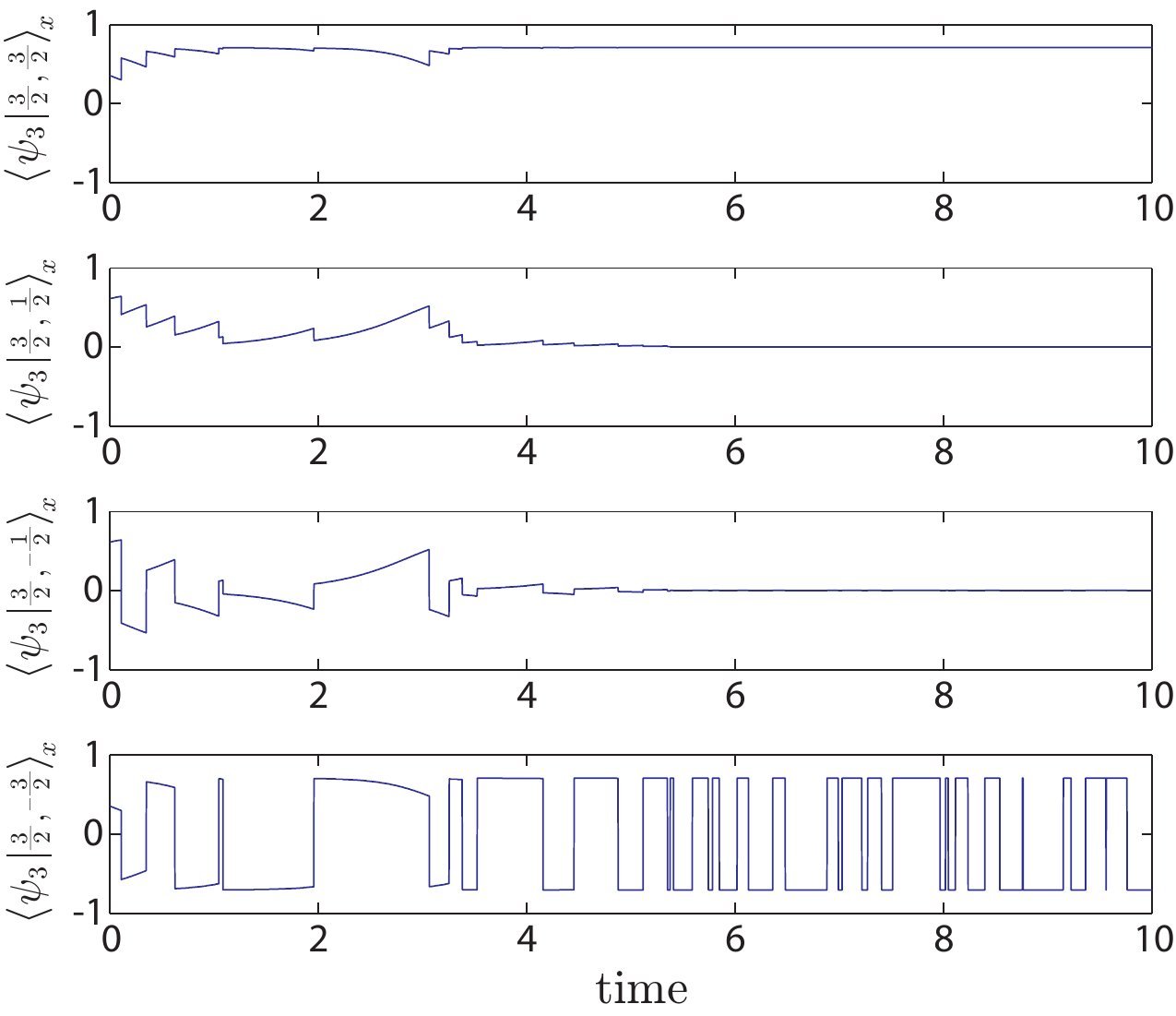}
\caption{Example trajectory for the initial state $\ket{3/2,-3/2}_{z}=\ket{000}_z$, where sufficient jumps occur over a time interval to promote the three-qubit state into a cycle involving $\ket{\chi^{\pm}_{3}(3/2)}$.}
\label{3QPic2}
\end{figure}

A trajectory in which the three-qubit state evolves to a cycle between the states $\ket{\chi^{\pm}_{3}(3/2)}$ (occurring with probability $1/4$) is shown in Fig.~\ref{3QPic2}. In this case, collapses are clearly much more frequent than in the previous example, since, mathematically, the norm of the wave function decays more rapidly between each collapse.  
In fact, the ratio of the frequencies of collapses for the two cycles is given by 
\begin{eqnarray}
\label{3Qjumps}
   \frac{\bra{\chi^{\pm}_{3}\!\left(\tfrac{3}{2}\right)}C^{\dag}C\ket{\chi^{\pm}_{3}\!\left(\tfrac{3}{2}\right)}}  
   {\bra{\chi^{\pm}_{3}\!\left(\tfrac{1}{2}\right)}C^{\dag}C\ket{\chi^{\pm}_{3}\!\left(\tfrac{1}{2}\right)}} = 9 \,,        
\end{eqnarray}                                                                                
so which cycle the system evolves into for any individual realization is clearly distinguishable from the photon counting record.

It is interesting to consider the states involved in the two cycles. In the $J_x$-eigenstate and $J_z$-eigenstate bases, respectively, we have
\begin{subequations}
\begin{eqnarray}
\label{3Qchi+bare1}
   \ket{\chi^{+}_{3}\!\left(\tfrac{1}{2}\right)} &=&  \frac{1}{\sqrt{6}}\left[\left(\ket{110}_x + \ket{101}_x + \ket{011}_x\right) \right. \nonumber
\\                                                     
   && ~~ + \left. \left(\ket{001}_x + \ket{010}_x + \ket{100}_x \right)\right] 
\\   
   &=& \frac{1}{2\sqrt{3}}\left( \ket{011}_z + \ket{101}_z + \ket{110}_z\right) \nonumber \\
                                                  && ~~ + \frac{\sqrt{3}}{2}\ket{000}_z \,, \\[0.2cm] 
\label{3Qchi-bare1} 
   \ket{\chi^{-}_{3}\!\left(\tfrac{1}{2}\right)} &=& \frac{1}{\sqrt{6}}\left[\left(\ket{110}_x + \ket{101}_x + \ket{011}_x\right) \right. \nonumber
\\                                                     
   && ~~ - \left. \left(\ket{001}_x + \ket{010}_x + \ket{100}_x \right)\right] 
\\   
   &=& \frac{1}{2\sqrt{3}}\left(\ket{100}_z + \ket{010}_z + \ket{001}_z\right) \nonumber \\
                                                  && ~~ - \frac{\sqrt{3}}{2}\ket{111}_z \,,
\end{eqnarray} 
\end{subequations} 
and
\begin{subequations}   
\begin{eqnarray}
\label{3Qchi+bare2}
   \ket{\chi^{+}_{3}\!\left(\tfrac{3}{2}\right)} &=& \frac{1}{\sqrt{2}} \left( \ket{111}_{x} + \ket{000}_{x} \right)
\\   
   &=& \frac{1}{2}\left(\ket{000}_z+\ket{011}_z+\ket{101}_z+\ket{110}_z\right) \,, \nonumber \\ \\ 
\label{3Qchi-bare2} 
   \ket{\chi^{-}_{3}\!\left(\tfrac{3}{2}\right)} &=& \frac{1}{\sqrt{2}} \left( \ket{111}_{x} - \ket{000}_{x} \right)
\\  
   &=& \frac{1}{2}\left(\ket{111}_z+\ket{100}_z+\ket{010}_z+\ket{001}_z\right) \,. \nonumber \\
\end{eqnarray}
\end{subequations}
In the $J_x$-eigenstate basis, the states $\ket{\chi^{\pm}_{3}(3/2)}$ are simply GHZ states,
while $\ket{\chi^{\pm}_{3}(1/2)}$ are each a linear combination of two $W$ states, $\ket{3/2,1/2}_x$ and $\ket{3/2,-1/2}_x$. Such combinations are sometimes referred to as $G$ states \cite{Sen&Zukowski}, which fall under the GHZ class of entangled states \cite{Dur00}, but are clearly less entangled than a single $W$ state.
Alternatively, in the $J_z$-eigenstate basis, the states $\ket{\chi^{\pm}_{3}(1/2)}$ and $\ket{\chi^{\pm}_{3}(3/2)}$ are all superpositions of a product state and a $W$ state \cite{Kiesel03, Eibl04}. 
Note again that unitary rotations of states between bases could be implemented using coherent Raman transitions between the atomic ground states $\ket{0}$ and $\ket{1}$.


\subsection{Entangled-state cycles for \textsl{N}=4}
\label{N=4}

In the case of four qubits with initial state $\ket{2,-2}_{z}=\ket{0000}_z$, (\ref{evo}) reads
\begin{eqnarray}
    |\bar{\psi}_{4}(t)\rangle &=& \frac{1}{\sqrt{2}} \:e^{-t} \ket{\chi^{+}_{4}(1)} + \frac{1}{\sqrt{8}} \:e^{-4t} \ket{\chi^{+}_{4}(2)} \nonumber \\ 
                       && + \:\sqrt{\frac{3}{8}}\ket{2,0}_x \,,
\end{eqnarray} 
so again two distinct entangled-state cycles are possible, corresponding to $m=1$ or $m=2$ and occurring with probabilities $1/2$ or $1/8$, respectively. In the $J_x$-eigenstate basis, the cycles occur again either between two GHZ states, $\ket{\chi^{\pm}_{4}(2)}$, or between two $G$ states, $\ket{\chi^{\pm}_{4}(1)}$.

\begin{figure}
\includegraphics[width=8cm]{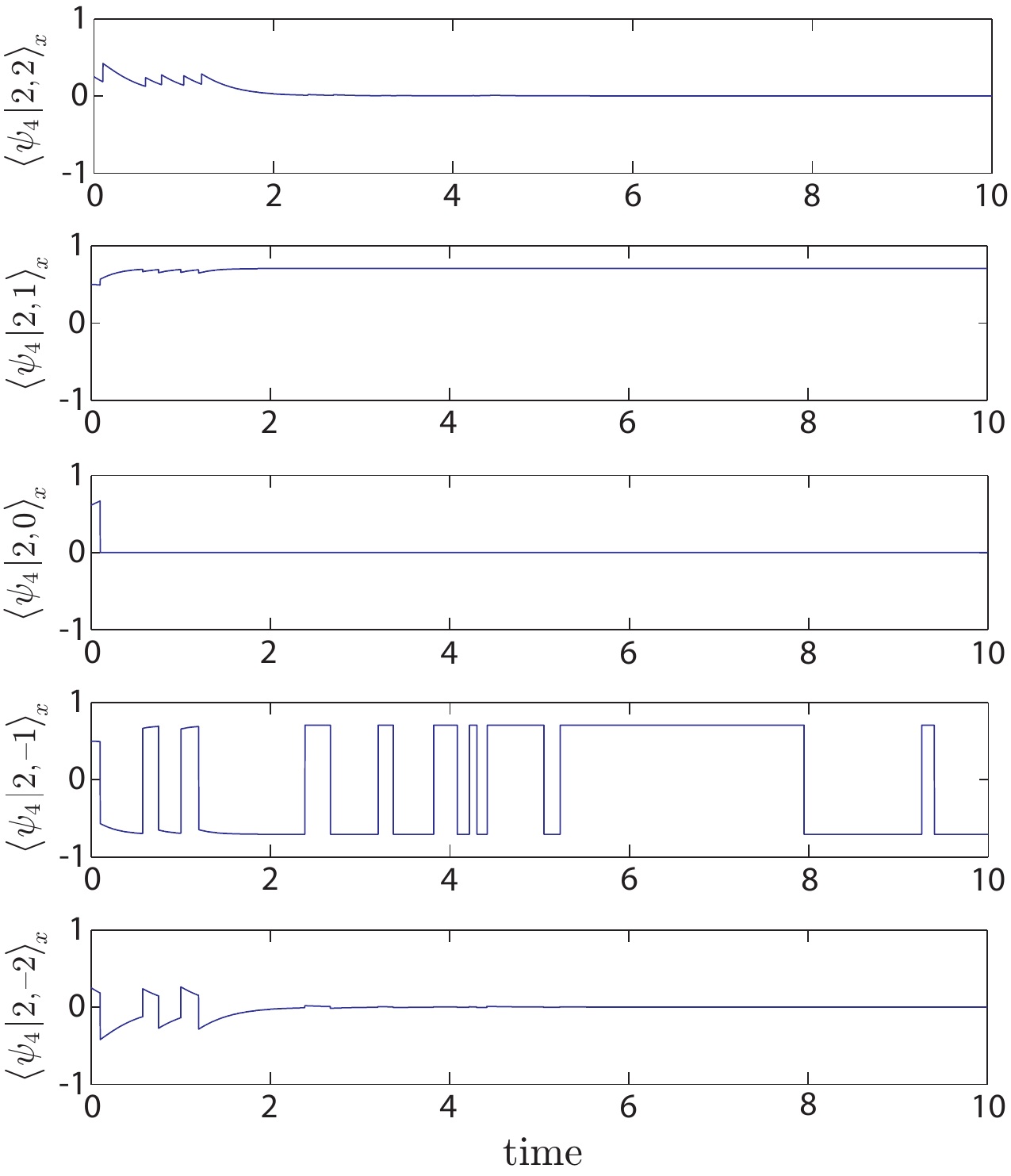} 
\caption{Example trajectory for $N=4$ with initial state $\ket{2,-2}_{z}=\ket{0000}_z$. In this case, the system settles into the cycle between the states $\ket{\chi^{\pm}_{4}(1)}$.}
\label{4QPic1}
\end{figure}

\begin{figure}
\includegraphics[width=8cm]{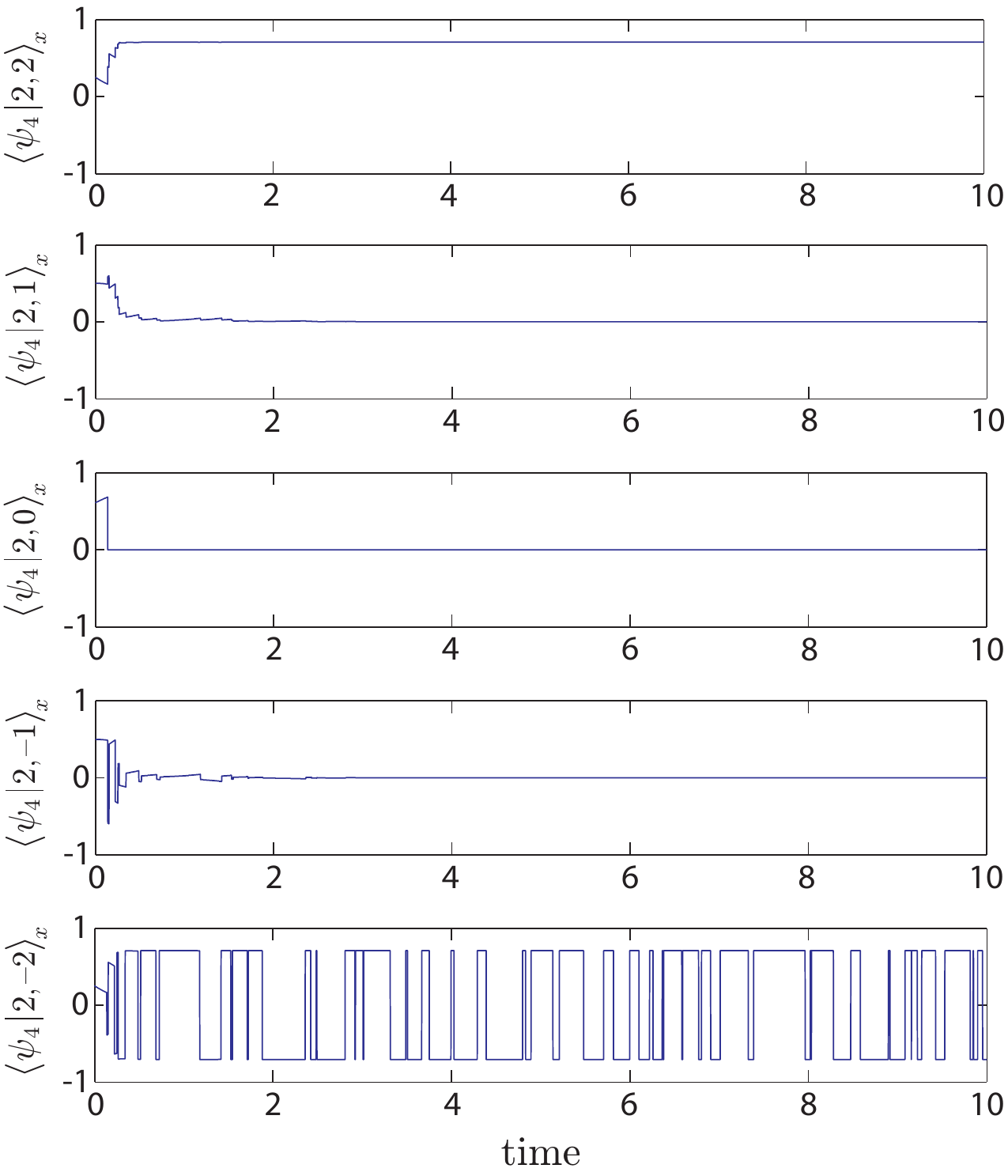} 
\caption{Example trajectory for $N=4$ with initial state $\ket{2,-2}_{z}=\ket{0000}_z$. In this case, the system settles into the cycle between the states $\ket{\chi^{\pm}_{4}(2)}$.}
\label{4QPic2}
\end{figure}

However, when represented in the $J_z$-eigenstate basis, the cycles now occur between quite different forms of entangled states. In particular, the $m=1$ cycle is between a GHZ state and a $G$ state, i.e.,  
\begin{subequations}
\begin{eqnarray}       
\label{4Qchi+bare1}   
   \ket{\chi^{+}_{4}(1)} &=& \frac{1}{\sqrt{2}}\left( \ket{0000}_z - \ket{1111}_z \right) \,, \\[0.2cm]
\label{4Qchi-bare1}   
   \ket{\chi^{-}_{4}(1)} &=& \frac{1}{\sqrt{8}}\left[ \left( \ket{0001}_z + \ket{0010}_z + \ket{0100}_z + \ket{1000}_z\right) \right. \nonumber \\ 
                          && - \left. \left(\ket{1110}_z + \ket{1101}_z + \ket{1011}_z + \ket{0111}_z \right) \right] \,, \nonumber \\
\end{eqnarray}
\end{subequations}
while the $m=2$ cycle is between a $G$ state ($\ket{\chi^{-}_{4}(2)}$) and a state that is 
an equal superposition of a GHZ state and a Dicke state with two excitations \cite{Toth07, Kiesel07}, i.e.,
\begin{subequations}
\begin{eqnarray}
\label{4Qchi+bare2}
   \ket{\chi^{+}_{4}(2)} &=& \frac{1}{\sqrt{8}} \left( \ket{1100}_z + \ket{1010}_z + \ket{1001}_z + \ket{0101}_z \right. \nonumber \\
                          && \left. + \ket{0011}_z + \ket{0110}_z + \ket{0000}_z + \ket{1111}_z \right) \,,\nonumber \\  \\[0.2cm]
\label{4Qchi-bare2}
   \ket{\chi^{-}_{4}(2)} &=& \frac{1}{\sqrt{8}} \left( \ket{0001}_z + \ket{0010}_z + \ket{0100}_z + \ket{1000}_z \right. \nonumber \\
                          && \left. + \ket{1110}_z + \ket{1101}_z + \ket{1011}_z + \ket{0111}_z \right) \,. \nonumber \\                         
\end{eqnarray}
\end{subequations}
(Note that the state $\ket{2,0}_x$ is of the same form as (\ref{4Qchi+bare2}), but with different weightings for the GHZ and Dicke states.) Examples of entangled-state cycles for $N=4$ are shown in Figs.~\ref{4QPic1} and \ref{4QPic2}.
%


\subsection{Larger $N$}

\begin{figure}
\includegraphics[width=8cm]{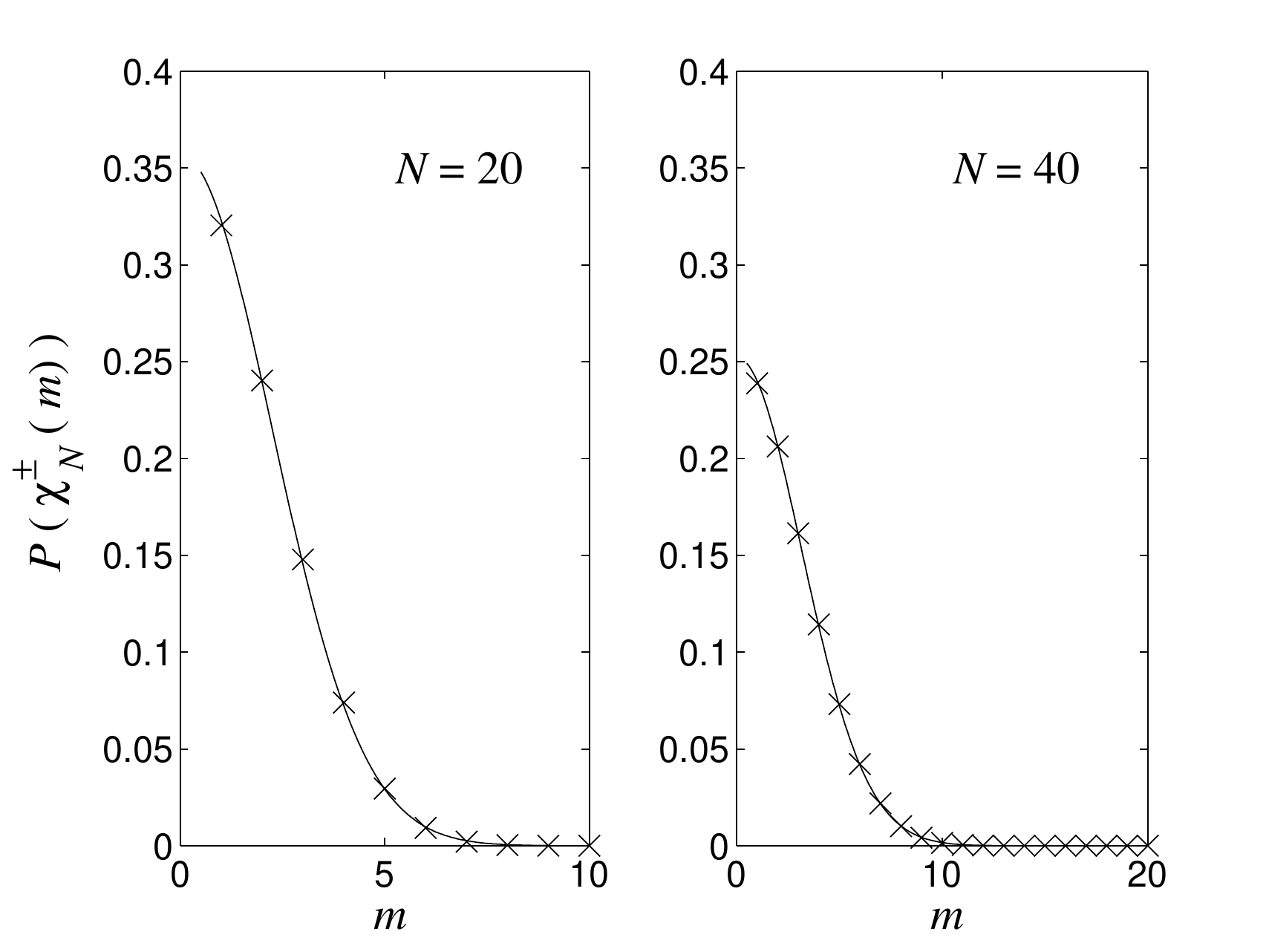} 
\caption{Probability $P(\chi_N^\pm (m))$ for the system to evolve into the entangled-state cycle between $\ket{\chi_N^\pm (m)}$. The crosses are calculated from Eq.~(\ref{Pchi}) The solid lines are the Gaussian approximation given in Eq.~(\ref{Pchi_Gauss}).}
\label{P_chi_m}
\end{figure}

For larger $N$ the number of possible cycles obviously increases, as does the variety of entangled
states involved. 
From the decomposition of the initial state $|0^{\otimes N}\rangle_{z}$ in terms of the states
$\ket{N/2,0}_x$ and $\ket{\chi_N^+(m)}$, one deduces that the probability, $P(\chi_N^\pm (m))$, 
for the system to be projected into the cycle involving the states $\ket{\chi_N^\pm (m)}$ is given by
\begin{eqnarray}
P(\chi_N^\pm (m)) &=& 
2\left| D^{(N/2)}_{m,-N/2}\left(-\frac{\pi}{2}\right) \right|^2 \nonumber
\\
&=& \frac{N!}{2^{N-1}(N/2+m)!(N/2-m)!} \, .
\label{Pchi}
\end{eqnarray}
In the large-$N$ limit, and for $m\ll N$, this is approximated well by the simple Gaussian 
form (derived using Stirling's approximation)
\begin{eqnarray}
P(\chi_N^\pm (m)) \simeq 2\sqrt{\frac{2}{\pi N}}\, \exp \left( - \frac{m^2}{N/2} \right) ,
\label{Pchi_Gauss}
\end{eqnarray}
as demonstrated in Fig.~\ref{P_chi_m} for $N=20$ and $N=40$. 
This result makes it very clear that cycles with $m>\sqrt{N}$ become increasingly unlikely for large
$N$. Favored cycles are those with small $m$, corresponding, in fact, to states satisfying the condition 
for genuine $N$-partite entanglement,
\begin{eqnarray}
\langle J_x^2 \rangle < \frac{N}{4} \, ,
\end{eqnarray}
i.e., states with $\langle J_x^2 \rangle =m^2 < N/4$.

%


\newpage

\section{Conclusion}
\label{conclusion}

In this work we have described a means of realizing, through suitable laser and cavity QED
interactions, a collective atomic spin system with dynamics described by the simple master 
equation
\begin{eqnarray*}  
   \frac{d\rho}{dt} = 2J_{x}\rho J_{x} - J^{2}_{x}\,\rho - \rho J^{2}_{x} .
\end{eqnarray*}     
Despite its apparent simplicity, this model reveals a remarkable diversity of behavior when 
viewed in a quantum trajectory picture, given an initial state 
$\ket{N/2,-N/2}_z = |0^{\otimes N}\rangle_{z}$ and assuming photon counting measurements 
on the cavity output field.
The different behaviors can be readily understood by employing a basis of $J_x$-eigenstates and
interpreting the configuration in terms of a projective measurement of the operator $J_x^2$.
On top of the interesting dynamical behavior, 
the actual states produced, either in steady state ($m=0$) or participating in an entangled-state 
cycle ($m>0$), typically possess genuine multipartite entanglement.

Our work is actually quite closely related, but complementary, to that of Stockton {\it et al}.  
\cite{Stockton04}, who consider continuous quantum nondemolition measurement of the 
atomic spin operator $J_z$, via a dispersive atom-cavity interaction and homodyne detection 
of the cavity field, starting from an initial atomic spin state polarized along the $x$-axis.
There, Dicke states in the $J_z$-eigenstate basis are prepared, either nondeterministically, with a 
probability determined by the initial state decomposition in this basis, or
deterministically, via the use of state-based feedback control. 
Such a feedback approach could perhaps also be applied usefully to the present 
configuration.


\begin{acknowledgments}
This work was supported in part by the Marsden Fund of the Royal Society of New Zealand. 
ASP thanks G. T\'oth and E. Solano for helpful discussions.
\end{acknowledgments}

\appendix

\section{Monte Carlo Wave-Function Routine}
\label{MCwavefuncn}

We will not discuss quantum trajectories in any profound way, but we simply outline how a realization of the reduced density operator for the system, called a trajectory, can be built operationally using a Monte Carlo algorithm. In essence, the trajectory formalism reduces the problem of mixed-state evolution to that of a pure state, as we describe below. References \cite{Carmichael89,Carmichael93,Dalibard92,Dum92} provide an excellent account of the formalism.
 
For our quantum trajectory approach, the output of the optical cavity is considered to be monitored continuously by photon counting, as shown in Fig.~\ref{PhysSys}(b). In this scenario, each photodetection sequence corresponds to a trajectory of the $N$-qubit collective atomic state.

Consider a normalized pure state $|\psi(t)\rangle$ at time $t$. The state is evolved ``freely'' over a small (ideally infinitesimal) time increment $dt$ according to
\begin{eqnarray}
\label{nonHermite}   
   |\psi(t+dt)\rangle = \frac{\left(1-i\,\mathcal{H}_{\rm eff}dt/\hbar\right)|\psi(t)\rangle}
                                   {\left\|\left(1-i\,\mathcal{H}_{\rm eff}dt/\hbar\right)|\psi(t)\rangle\right\|} ,                 
\end{eqnarray}
where $\mathcal{H}_{\rm eff}$ is a non-Hermitian effective Hamiltonian and the state is renormalized after the non-unitary evolution \cite{Note1}. Such free evolution is interspersed, at random times, with ``quantum jumps'' or ``collapses'', the probability of which must be calculated at each timestep. In particular, the probability that a collapse of the state occurs over the (infinitesimal) time step $dt$ is given by
\begin{eqnarray}
\label{CollapseOp}
   \mathcal{P}(t) = \langle\psi(t)|C^{\dag}C|\psi(t)\rangle\,dt \,,
\end{eqnarray}  
where $C$ is the collapse operator.
To determine whether a collapse occurs or not, a uniformly distributed random number $u(t)$ is drawn from the unit interval $[0,1]$ and compared to $\mathcal{P}(t)$. If $\mathcal{P}(t)>u(t)$ then the state becomes 
\begin{eqnarray}
\label{CollapsedState}
   |\psi(t+dt)\rangle = \frac{C|\psi(t)\rangle}{\|C|\psi(t)\rangle\|}\, ,
\end{eqnarray} 
otherwise, for $\mathcal{P}(t) \le u(t)$, the state is not collapsed and it continues to evolve, according to (\ref{nonHermite}), over another subsequent interval of $dt$. This process is iterated over sufficiently many time steps to build up a trajectory. Note that for our system the action of $C$ represents the emission and subsequent detection of a photon in the cavity output field.



\begin{thebibliography}{100}
\bibitem{multipartiteE}
As defined in \cite{Toth07}, an $N$-qubit state is said to possess genuine multipartite
entanglement if it is not biseparable, i.e., if it cannot be expressed in the form
$|\Phi_{1\ldots m}\rangle\otimes |\Phi_{m+1\ldots N}\rangle$, which is separable
with respect to the partition $(1,2,\ldots,m)(m+1,m+2,\ldots,N)$.
\end{thebibliography}
\end{document}